\documentstyle[aps,preprint]{revtex}
\begin{document}
\tightenlines

\title{The Pair Contact Process in Two Dimensions}
\author{Jafferson Kamphorst Leal da Silva$^{\dagger}$
and
Ronald Dickman$^{\ddagger,a}$}
\address{
Departmento de F\'\i sica,
Universidade Federal de Minas Gerais, Caixa Postal 702,
30161-970, Belo Horizonte - MG, Brasil}
\date{\today}
\maketitle
\begin{abstract}
We study the stationary properties of the 
two-dimensional pair contact process, a nonequilibrium
lattice model exhibiting a phase transition to an absorbing state
with an infinite number of configurations.
The critical probability and static critical exponents are
determined via Monte Carlo simulations, as well as order-parameter
moment ratios and the scaling of the initial density decay.
The static critical properties are consistent with the
directed percolation universality class.
\vspace {0.3truecm}

\noindent PACS numbers: 05.50.+q, 02.50.-r, 05.70.Ln
\end{abstract}
\vspace{1.0truecm}

\noindent 
$^{\dagger}${\small electronic address: jaff@fisica.ufmg.br }\\
$^{\ddagger}${\small electronic address: dickman@fisica.ufmg.br } \\
$^a${\small On leave of absence from: Department of Physics and Astronomy,
Herbert H. Lehman College, City University of New York,
Bronx, NY, 10468-1589.} \\
 
\newpage

\section{Introduction}

Critical phenomena at absorbing-state phase transitions (i.e., between an
active state and one 
in which the dynamics is frozen), are of longstanding interest in statistical 
physics, and have enjoyed renewed attention due to connections with 
epidemics \cite{harris},
catalytic kinetics \cite{zgb,evans}, surface growth \cite{alon},
self-organized criticality \cite{gz,maslov,dvz,vdmz}, and issues of scaling and 
universality \cite{privbook,marro}.  In these systems, conflict between two
opposing processes (e.g., creation and annihilation), typically
leads to a continuous transition at a critical parameter value. 
Such transitions are known to fall 
generically in the universality class of directed percolation (DP) 
\cite{janssen,gr1,glb}, although the critical behavior is modified in
the presence of local parity conservation 
\cite{gkt,baw1,iwanbaw,cardybaw}.  

Another interesting case 
(without a conservation law), appears when the dynamics
can become trapped in one of an infinite
number (in the thermodynamic limit) of absorbing configurations (INAC).
Systems of this sort were introduced in catalysis modeling 
\cite{benav,albdd,yaldram}; their critical properties have been studied 
in detail 
by various workers \cite{ijnoco,pcp1,pcp2,mendes,inas,snr,mgd,gbr}.
In one dimension, the pair contact process (PCP) \cite{pcp1}, and other 
models with INAC exhibit static critical behavior in the DP class 
\cite{pcp2,rdjaff}, but the critical exponents associated with the spread
of activity from a localized seed are {\it nonuniversal}, varying 
continuously with the particle density in the environment \cite{pcp2,gbr},  
and follow a generalized hyperscaling relation \cite{mendes,mgt}.
The anomalous spreading can be
traced to a long memory in the dynamics of the order parameter, $\rho$,
arising from coupling 
to an auxiliary field that remains frozen in regions where
$\rho = 0$ \cite{inas,mgd,gbr}.  A field theory (i.e., a stochastic partial            
differential equation for $\rho({\bf x},t)$), incorporating this
memory term reproduces the nonuniversal exponents observed in
simulations \cite{lopez}.

In two dimensions the situation is much less clear.
Simulation results for a microscopic model with INAC \cite{snr} conflict
with studies of models exhibiting the aforementioned 
long memory \cite{mgd,gbr}.
(In particular, it seems possible that for one range of densities 
the spreading dynamics is that of dynamic percolation, while for
another range there is compact growth, perhaps with nonuniversal
exponents \cite{mgd}, or without well defined scaling behavior \cite{gbr}.)
In hopes clarifying the nature of critical spreading in the
presence of INAC, we propose to study the pair contact process in two
dimensions.  Our interest in the PCP is motivated by its simplicity, compared
with the model studied in Ref. \cite{snr}.
The present work is devoted to static critical behavior, and provides 
the critical parameter value
and the ``natural density" in the absorbing state (defined below) needed 
for a detailed study of spreading.
These results confirm that the static behavior falls in the DP
class; analyses of moment ratios and the initial decay of the order parameter
provide further support.  The balance of this paper is devoted to
defining the model and simulation algorithm (Sec. II), simulation
results (Sec. III), and a brief summary (Sec. IV).

\section{Model}

The pair contact process (PCP) is an
{\em interacting particle system}: a Markov process
whose state space is a set of particle configurations
on a lattice \cite{liggett,konno}.
Each nearest-neighbor pair of particles has a rate $p$
of mutual annihilation, and a rate $1-p$ of attempted creation.
In a creation attempt on the square lattice, a new
particle may appear (with equal likelihood) at any of the six sites
neighboring the pair, provided the chosen site is vacant.
(Attempts to place a new particle at an occupied site fail.)
The PCP exhibits an active phase for $p < p_c$;
above this value the system falls into an
absorbing configuration that typically contains a substantial
density, $\phi$, of particles. (Any arrangement of particles devoid
of nearest-neighbor pairs is absorbing.)  

In our simulations, we maintain a list of the $N_p$ current nearest-neighbor
pairs.  At each step we choose a pair at random from the list, and a
process (annihilation with probability $p$, creation with probability $1-p$).
In case of annihilation, the two particles are simply removed.  For creation, 
we choose a site {\bf x} at random from among the six neighbors of the pair, 
and place a new particle there if {\bf x} is currently vacant.  
(If {\bf x} is occupied the configuration remains the same.)  
The time increment associated with this step is $\Delta t = 1/N_p$,
corresponding to one transition per pair per unit time, in agreement with 
the transition rates that define the process.
Following each change we update the list of pairs.
We use a square lattice of $L \times L$ sites, with periodic boundaries;
in the studies reported here, all sites are initially occupied.

\section{Critical properties}

To locate the critical point $p_c$ we study the size-dependence of the 
(quasi) stationary
pair density $\rho$, i.e., the fraction of nearest neighbors harboring 
a pair of particles,
in surviving trials, following a transient during which 
$\rho (t)$ relaxes from its initial value of
unity.  $\rho$ is the order parameter for the PCP, and as such
we expect that at the critical point,
\begin{equation}
\rho(p_c,L) \sim L^{-\beta/\nu_{\perp}} \;,
\label{rhosc}
\end{equation}
while off-critical values of $p$ should yield deviations from the power law.
We studied
the pair density in systems of size $L =$ 10, 20, 40, 80 and 160, 
for times $t_m$ ranging
from $10^3$ for $L=10$ to $5 \times 10^4$ for $L=160$, with sample sizes ranging from
$10^6$ trials ($L=10$) to $10^4$ trials ($L=160$).  The results (see Fig. 1) show
$\rho (p,L)$ following a power law for $p=0.2005$, but clearly not for $p=0.200$ or
0.201.  We conclude that $p_c = 0.2005(2)$, the figure in parentheses denoting the
uncertainty of our estimate.  The data for $p=0.2005$ yield the exponent ratio
$\beta/\nu_{\perp} = 0.793(5)$, in good agreement with the value of 0.799(2)
for DP in 2+1 dimensions \cite{rew}.  

We also determined the survival probability, $P(t,p,L)$, i.e., the probability
that the system contains at least one nearest-neighbor pair.  For finite $L$, 
this decays asymptotically as $P \sim e^{-t/\tau_P}$, with
the lifetime showing a power-law dependence on the system size at
the critical point:
\begin{equation}
\tau_P(p_c,L) \sim L^{\nu_{||}/\nu_{\perp}} \;.
\label{tausc}
\end{equation}
The data for $p=0.2005$ (see Fig. 2) yield $\nu_{||}/\nu_{\perp} = 1.79(1)$, 
reasonably close to the DP value, 1.766(2).  We find that $\rho$ and the particle
density $\phi$  
also approach their stationary values exponentially: 
$\rho (t) - \rho_s \sim e^{-t/\tau}$ (similarly for $\phi$), but on
a much shorter time scale than that of $P(t)$: $\tau \simeq \tau_P/10$.
Analysis of the data for $\tau$ at the critical point yields
$\nu_{||}/\nu_{\perp} = 1.69(3)$.

For $p \leq p_c$ the process always falls into the absorbing state.  
The properties of this
state are determined by the
probability distribution (induced by the dynamics) on the set of absorbing 
configurations for system size $L$.
Of interest is the particle density $\phi$ in the absorbing state, 
in particular, the ``natural"
density, defined as the limiting value at the critical point:
\begin{equation}
\phi_{nat} \equiv \lim_{L \rightarrow \infty} \phi(p_c,L) .
\label{phinat}
\end{equation}
(In one dimension, it is only for this particle density that the spreading 
exponents take DP values \cite{pcp2}.)   In our simulations at $p_c$ virtually
all of the trials end in an absorbing configuration before $t_m$; the final
particle density in the absorbing state yields an estimate for 
$\phi(p_c,L)$.  

Since the dynamics of $\phi$ is tied to that of the order parameter, $\rho$,
and since the excess particle density ($\phi(p) - \phi_{nat}$) in a related one-dimensional
model is known to be governed by the order-parameter exponent 
$\beta$ \cite{pcp2}, we expect that the leading finite-size correction
to the particle density to be $\sim L^{-\beta/\nu_{\perp}}$, just as for $\rho$. 
This is confirmed in Fig. 3. Linear fits to the
data for $L \geq 40$ yield 
$\phi \simeq 0.1480 + a L^{-\beta/\nu_{\perp}}$ and
$\rho \simeq b L^{-\beta/\nu_{\perp}} $, (with $a = 0.9662$ and $b=1.426$),
suggesting that the linear combination $\phi - (a/b) \rho$ will
be essentially independent of $L$.  This is indeed so for $L \geq 20$,
as shown in the inset of Fig. 3, from which we obtain our final
estimate, $\phi = 0.1477(1)$.

Order parameter moment ratios provide another tool for assigning a model
a universality class; in equilibrium spin systems Binder's reduced fourth
cumulant has been widely used for this purpose \cite{binder}.  A variety
of ratios, involving both odd and even moments, have been determined
for several one dimensional models with absorbing-state 
transitions (including the PCP), as well as for the two-dimensional contact process \cite{rdjaff}.
We determined the stationary order-parameter moments $m_1$,...$m_4$
($m_j \equiv \langle \rho^j \rangle$) in order to evaluate various ratios; 
the results are listed in Table I.  In Fig. 4
we plot several of the moment ratios versus $L^{-1}$; linear fits yield
the infinite-$L$ estimates given in Table I.  The latter agree quite
well with the results for the contact process, providing further support
for the PCP belonging to the DP universality class.  Curiously, the
moment ratios for the PCP appear to approach their limiting values
monotonically (with $L$), while the two-dimensional CP exhibits a
nonmonotonic $L$-dependence (see Fig. 5 of Ref. \cite{rdjaff}).

Finally, we analyzed the initial decay of the pair density at the critical point.
In general, we expect the order parameter to decay as a power law,
$\rho \sim t^{-\delta}$ in a critical system at short times
(i.e., $t < \tau \sim L^{\nu_{||}/\nu_{\perp}}$, for which the
correlation length $\xi < L$).  Our results,
plotted in Fig. 5, show a power-law decay with an exponent of 0.443(5).
This is somewhat smaller than, but still consistent with, recent estimates
of $\delta $ for DP in 2+1 dimensions, which range from
0.4505(10) \cite{vz} to 0.452(1) \cite{rew}.

\section{Summary}

We studied the stationary critical properties of the pair contact
process in two dimensions.  On the basis of the exponent ratios
$\beta/\nu_{\perp}$ and $\nu_{||}/\nu_{\perp}$, moment ratios, and
the initial decay of the order parameter, we can assign the PCP to
the directed percolation universality class, generic for absorbing-state transitions
without a conservation law or special symmetry.  We have noted several
minor discrepancies between our results and the standard DP values,
but expect that these are due to finite-size effects and/or a small
error in $p_c$, and do not reflect non-DP universality.  The issue of
spreading dynamics will be addressed in future work. 

\newpage

JKLS acknowledges partial support by CNPq.

\newpage

\begin{table}
\caption{Ratios of order-parameter moments in the critical PCP.
Entries for $L = \infty$ represent linear extrapolations;
data for the CP from Ref. [29].
Numbers in parentheses denote 
uncertainties in the last figure.}
\begin{center}
\begin{tabular}{|r|l|l|l|l|} 
$L$    &  $m_2/m_1^2$ & $ m_3/m_1^3 $ &$m_3/(m_1 m_2)$ & $m_4/m_2^2$ \\
\hline\hline
$20$  &   1.362(3)     &  2.300(6) &   1.638(5) &  2.247(10)    \\
$40$  &   1.343(3)     &  2.147(7) &   1.599(5) &  2.159(9)      \\
$80$  &   1.334(3)     &  2.111(9) &   1.582(6) &  2.116(10)    \\
$160$  & 1.327(4)     &  2.086(9) &   1.571(8) &  2.093(13)    \\
\hline
$\infty$ & 1.323(3)    &  2.067(9) &    1.56(1) & 2.07(1)        \\
\hline
CP      &  1.326(1)    &  2.080(1) &    1.569(1) & 2.093(8)
\end{tabular}
\end{center}
\label{pcprat}
\end{table}

\newpage

\noindent {\bf Figure Captions}
\vspace{1em}

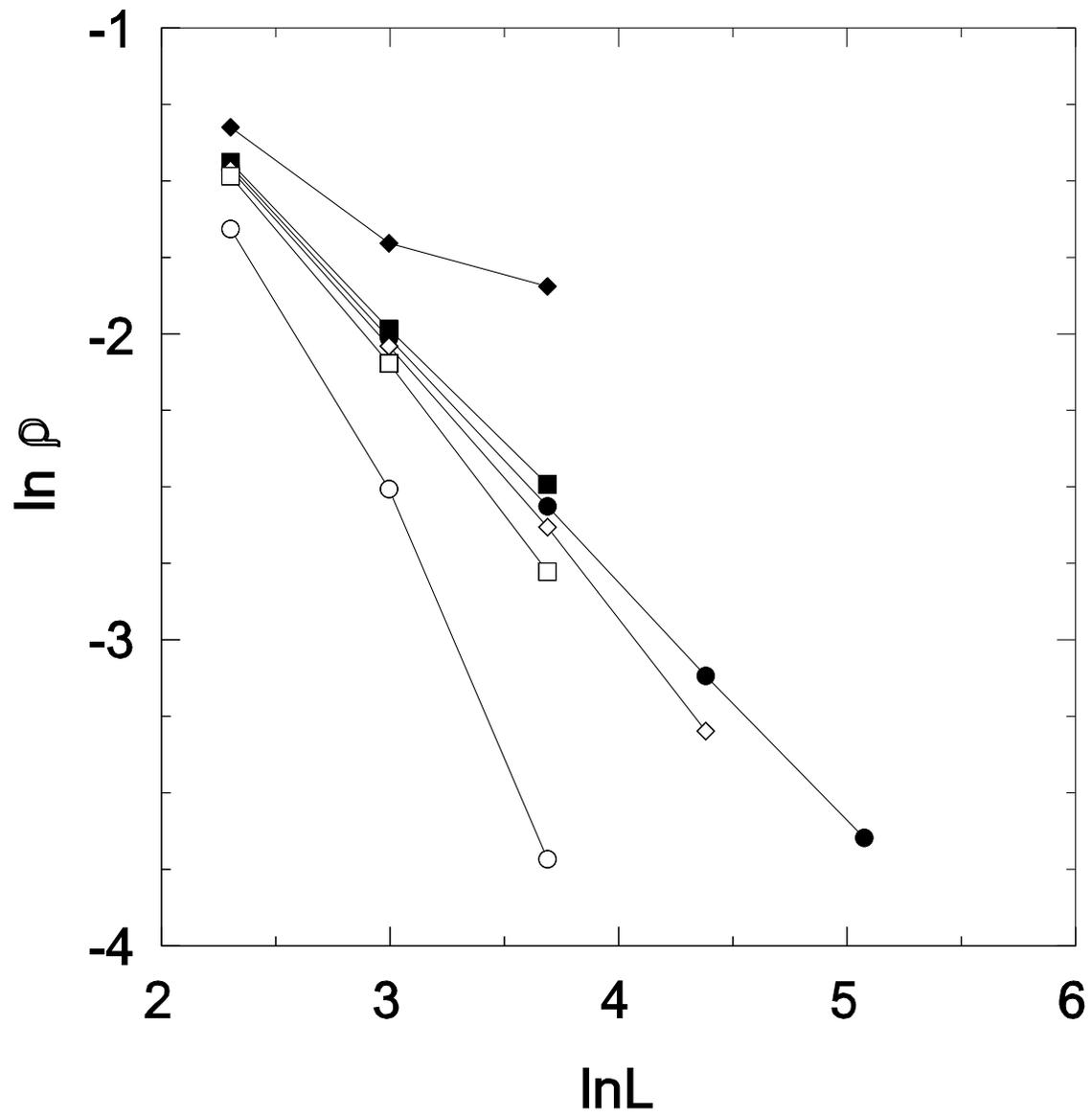
\noindent FIG. 1. Stationary pair density versus system size for $p$ =
0.21, 0.202, 0.201, 0.2005, 0.200, and 0.195 (left to right).
\vspace{1em}

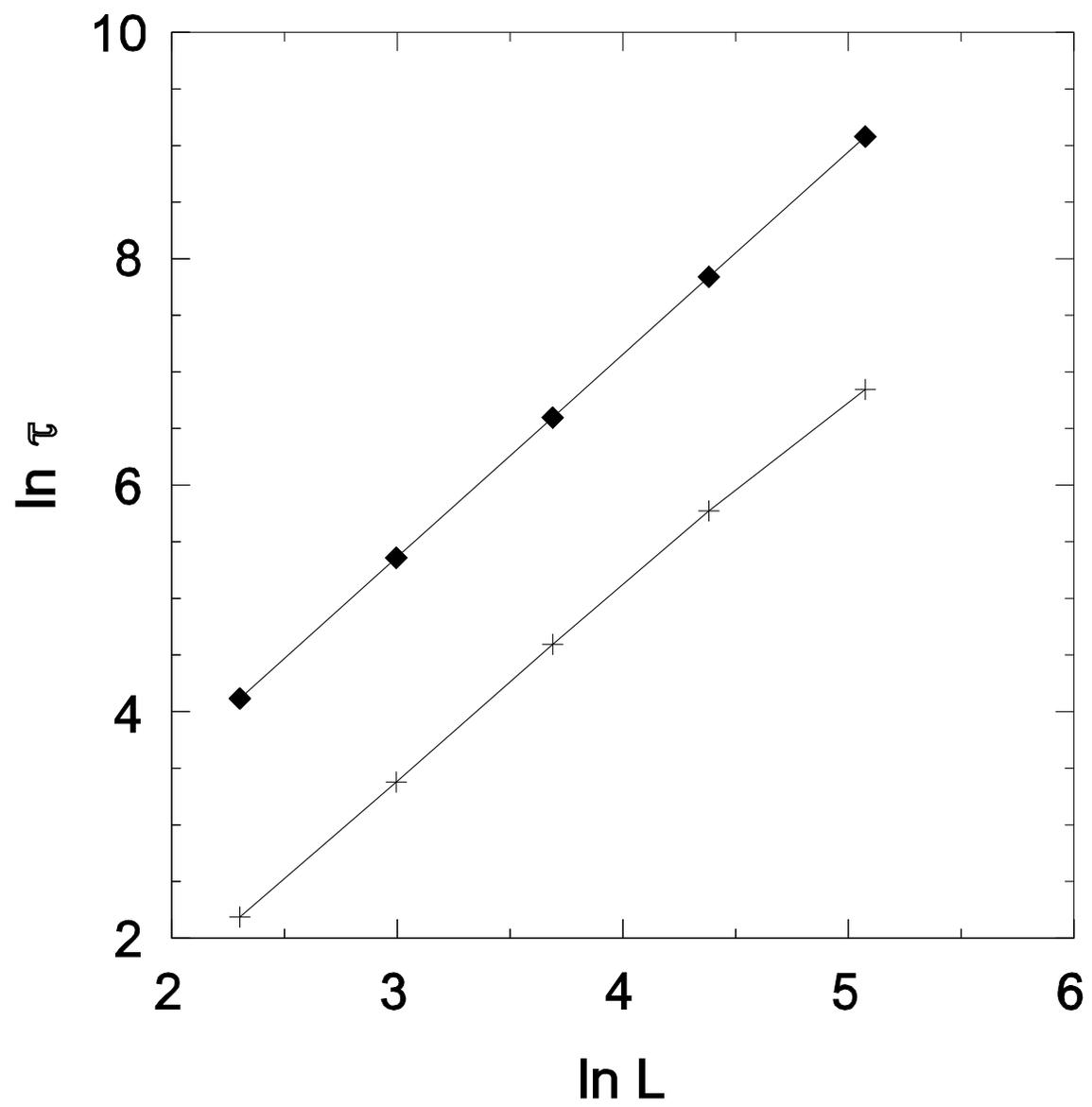
\noindent FIG. 2. Relaxation times versus system size at the critical point.
Upper set: $\tau_P$, the mean lifetime; lower set: $\tau$, associated with the relaxation of
$\rho$ and $\phi$.
\vspace{1em}

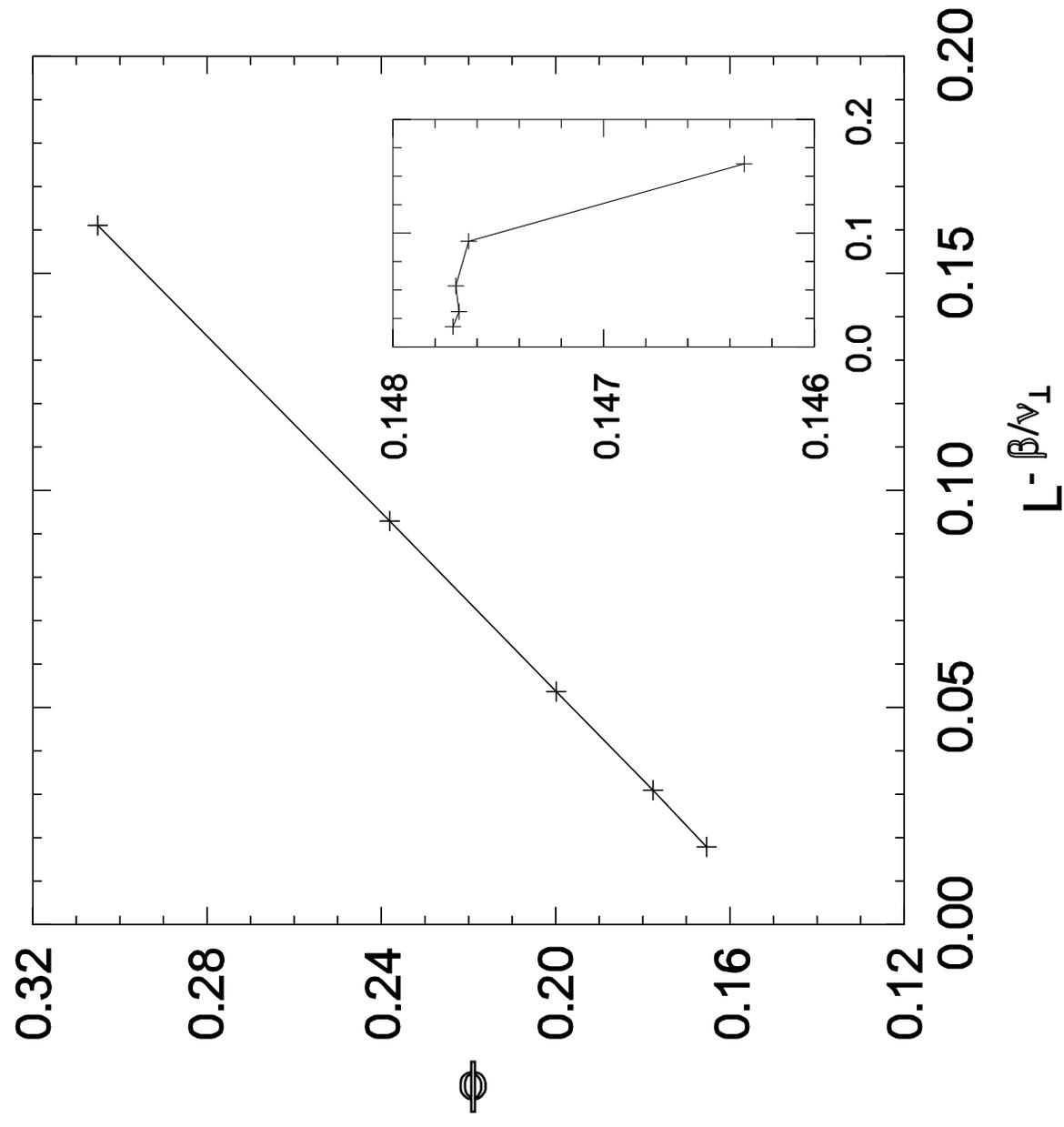
\noindent FIG. 3. Particle density in absorbing configurations at the critical point
versus $L^{-\beta/\nu_{\perp}} $.  The inset is a plot of $\phi - 0.6776 \rho$.
\vspace{1em}

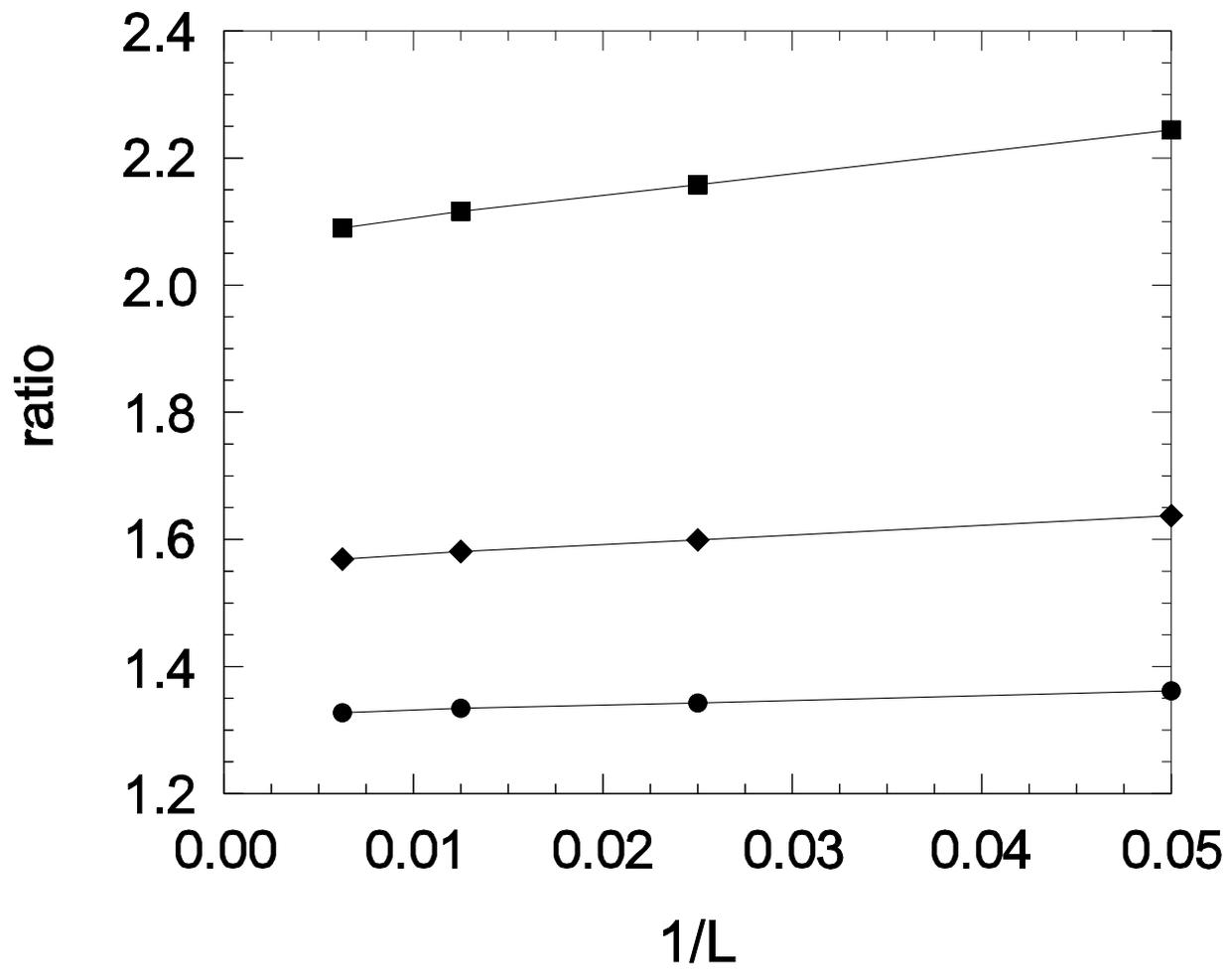
\noindent FIG. 4. Order-parameter moment ratios in the critical PCP.
Upper set: $m_4/m_2^2$; middle: $m_3/(m_1 m_2)$; lower: $m_2/m_1^2$.
\vspace{1em}

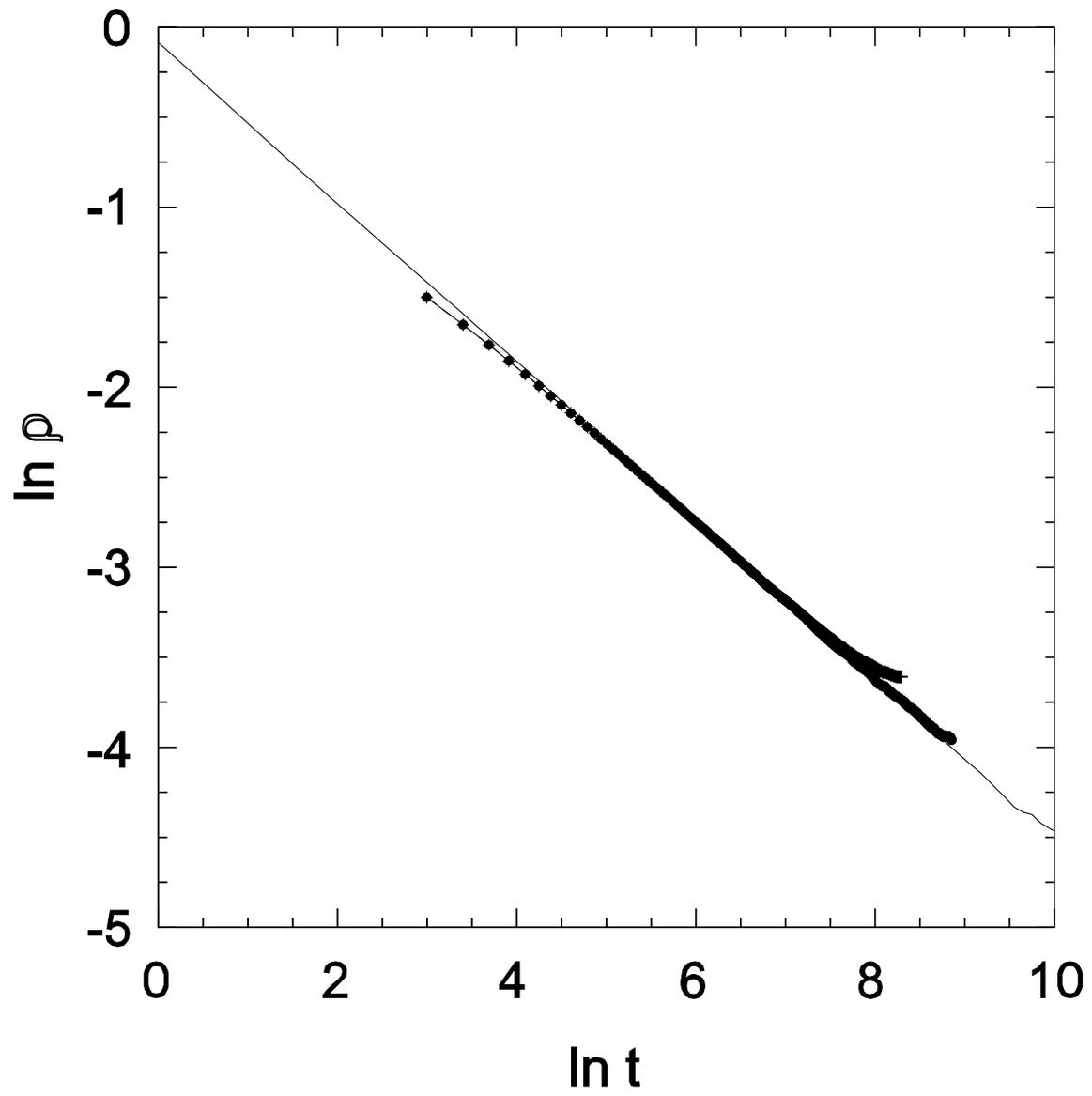
\noindent FIG. 5. Decay of the order parameter in the critical PCP.
$+$: $L=160$; $\bullet$: $L=320$; line: $L=640$.
\end{document}